\documentclass[preprint2]{aastex}
\usepackage{natbib}
\usepackage{graphicx}

\received{}
\revised{}
\accepted{}

\ccc{}
\cpright{}

\slugcomment{Submitted to ApJ}

\shorttitle{Henize 2-10: fresh view with NACO}
\shortauthors{Cabanac, Vanzi, Sauvage}

\begin{document}

\title{A fresh view on Henize 2-10 with VLT/NAOS-CONICA\footnote{based on observations made at the ESO 
VLT under program 71.B-0492, and 270.B-5011}}
\author{R\'emi A. Cabanac\footnote{also at Dep. de Astronom\'{\i}a y 
Astrof\'{\i}sica, Pontificia Universidad Cat\'{o}lica de Chile, 
Campus San Joaqu\'{\i}n, Vicu\~{n}a Mackenna 4860, Casilla 306, Santiago 22, 
Chile, and Visiting scientist at ESO, Santiago}}
\affil{Canada-France-Hawaii Telescope}
\affil{65-1238 Mamalahoa Highway, Kamuela, HI 96743, USA}
\email{cabanac@cfht.hawaii.edu}
\author{Leonardo Vanzi}
\affil{European Southern Observatory}
\affil{Alonso de Cordova, 3107, casilla 19001, Vitacura, Santiago, Chile}
\and \author{Marc Sauvage}
\affil{CEA/DSM/DAPNIA/SAp - UMR AIM}
\affil{CE Saclay, 91191 Gif sur Yvette CEDEX, France}
\begin{abstract}
We present high-resolution observations of Henize 2-10 in $K_S$ ($2.2\,\mu$m),
$L'$ (3.8$\,\mu$m), $M'$ (4.8$\,\mu$m) bands.  These allow for the first time 
to track accurately the structures at the heart of the galaxy from the optical 
to the radio. All radio knots previously observed can now be associated 
with $L'$ and $K_S$ emitting regions. This implies a revision of their 
physical nature. Instead of highly extinguished ultra-dense HII regions, we 
propose that two of the 5 radio knots are either supernova remnants or 
{\em normal} HII regions, while the remaining three are bona fide ultra-dense
 HII regions, although less obscured than was previously thought.
\end{abstract}

\section{Introduction}\label{intro}

Henize 2-10 (\objectname[He2-10]{He\,2-10}) is a blue compact galaxy with 
quite interesting properties. 
It is a nearby starburst galaxy with a heliocentric distance of 
$9\pm5$ Mpc \citep{johansson87,tully88}. In this paper we adopt a distance 
of 9 $h^{-1}$Mpc, with $H_0=75$ km/s, which yields a scale of 
$\sim$45pc/arcsec. The galaxy shows spectroscopic Wolf-Rayet features 
which indicates the presence of a very young starburst region
\citep{hutsemekers84,vacca92}, and has a slightly sub-solar metallicity 
\citep{kobulnicky99b}.
Figure~\ref{figKs} shows a RGB-composite image (in color in electronic version 
only) of the central $18\arcsec\times27\arcsec$ ($800\times1200$\,parsecs) 
of He\,2-10, where most 
of the current star formation occur. The blue and green channels are 
0.3\arcsec-convolved HST WFPC archive in $F555W$ and $F814W$. The red channel 
is our $0.3\arcsec$-seeing $K_S$ taken with VLT/ISAAC. The bright central
nucleus, generally referred to as region A, is an arc of UV-bright super-star 
clusters \citep{vacca92}. It is surrounded by two presumably older 
star-forming regions. Region B on the east shows 
a mixed population of blue and red clusters (only detected in $K_S$ 
and longer wavelength bands). Region C, on the north-west side, has a long 
tail containing bright red clusters as well. 
Dust is clearly apparent in this image, as shown by the red filaments 
observed to the south-east side between A and B. These appear
to be blown away from the central region, rapidly dissolving in He\,2-10's
interstellar medium. Assuming for the dust clouds a velocity equal to the 
typical sound speed in the ISM, i.e. 10 
$\sim 10$ km$\cdot$s$^{-1}$, hence a growing rate of $\sim10$pc$/$Myr, 
the present radii of curvature of the filaments $\sim 50-100$\,pc yield 
dynamical ages of $~5-10$Myr.  
Region A is also flanked by two compact red sources that 
are only visible at $K_S$ and longer wavelengths. 
These two sources get brighter with longer 
wavelengths, as shown in Figure~\ref{figmulti}. The various colors of the 
sources hint at a highly heterogeneous dust content, and possibly age 
differences among the cluster population.

Recent observations in the optical, IR, and radio, have brought new exciting 
facts. 

First, the youth of the starburst event in the center was confirmed by STIS 
analysis of the brightest UV/optical knots by \citet{chandar03} which yielded 
a coeval formation age of 4-5 Myr for all optical clusters.

Second, the presence and importance of dust in the central region was 
confirmed by high-resolution mid-infrared observations 
\citep{sauvage97,beck01,vacca02}: the majority (60\%) of the MIR emission 
is confined to a $\sim5"$ region, compatible in size with the location of 
the observed starburst. However the intrinsically large uncertainties in MIR 
astrometry prevented a clear identification of the MIR sources. 

Finally, the radio observations of \citet{kobulnicky99,kobulnicky00} 
evidenced 5 compact radio sources (hereafter called the radio knots) 
characterized by mostly thermal spectra. The striking morphological 
resemblance between the radio knots and the MIR emission allowed to tie 
down the location of the MIR sources precisely \citep{beck01}. 
Furthermore, comparing with HST images, \citet{kobulnicky99,kobulnicky00} 
argued that most of these radio-MIR sources were off-centered, in the 
dusty area between region A and B. They thus attributed this emission to 
young ($1<$Myr) ultra dense (UD) HII regions with ongoing star formation 
hidden in dense molecular clouds. Following this interpretation, 
\citet{beck01}, from $11.7\mu$m observations, derived that up to 
$10^4$ O7V stars (equivalent to $10^{49}$ Lyman photons$\cdot$s$^{-1}$) 
must be hidden in these dense cocoons, and \citet{vacca02}, 
from $10.8\,\mu$m observations, computed that $10^7 M_\odot$ of dust and 
gas must be surrounding the UD HII regions and estimated the bolometric 
luminosity of the brightest MIR region overlapping with radio knot 4 to be 
as much as $2\times10^9 L_\odot$. \citet{johnson03} extended the radio 
observations and refined the measurement of physical properties 
of the UD HII regions, in particular an anomalous low mass HII content 
of 2-8$\times10^3 M_\odot$ was found, attributed to the extreme youth 
of the objects. 

Therefore, as of today, He\,2-10 appears as a spectacular case of a starburst 
galaxy where a large fraction of its most current star formation activity 
lies completely buried in dust, and has absolutely no visible counterpart.

We present here high-resolution observations in $K_S$ ($2.2\mu$m) 
with VLT/ISAAC (\dataset[270.B-5011(A)]{under ESO program 270.B-5011(A)}), 
$L'$ (3.8$\mu$m), and $M'$ (4.8$\mu$m) bands with the Adaptive Optics 
VLT\,/\,NAOS\,-\,CONICA (\dataset[71.B-0492(A)]{under ESO program 71.B-0492(A)}) 
that give the highest resolution to date of the nucleus of He\,2-10 
in the NIR, a wavelength regime adequately located between the 
stellar optical regime and the dust thermal regime. The high quality of the 
observations allows the identification, for the first time, of bright $L'$ 
regions which correlate with radio knots \citep{kobulnicky99}, and $K_S$ 
regions that correlate with the optically bright cluster, thus bridging the 
existing wavelength gap.

\section{Observations and data reduction}\label{observations}


\begin{figure*}
\begin{center}
\includegraphics[width=\textwidth]{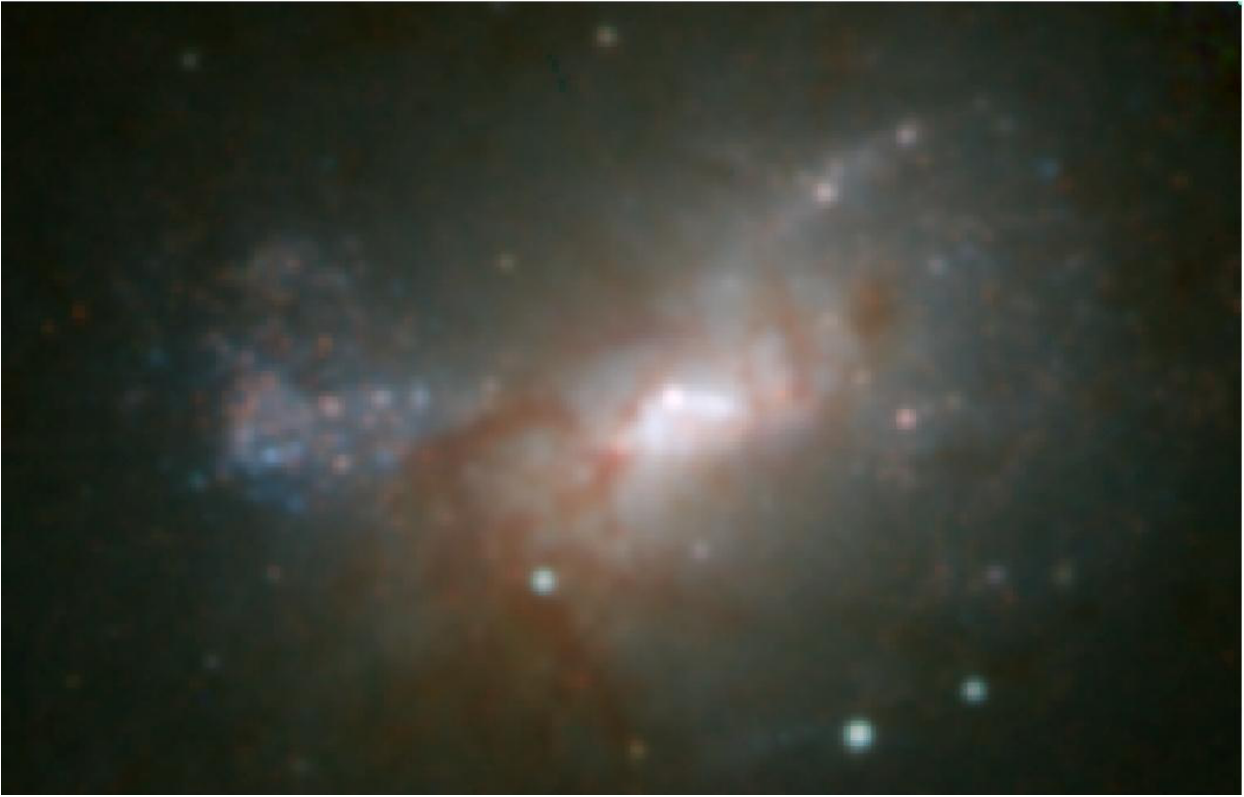}
\caption{\footnotesize He\,2-10 is shown in a RGB composite image (in color
in the electronic version only) of 
$18\arcsec\times27\arcsec$ ($800\times 1200$\,pc at a distance of 9\,Mpc). 
North is up and East is left. The blue and green 
channels are $0.3\arcsec$-convolved HST WFPC archive images in $F555W$ and 
$F814W$. The red channel is a $0.3\arcsec$-seeing $K_S$ image taken with 
VLT/ISAAC. The red filaments correspond to absorption regions in HST images, 
hence appear red by contrast. The region on the left shows a mix of bluish 
clusters and red clusters, hinting at a highly heterogeneous dust content, 
and possibly age differences. The bright red sources on each side of the 
nucleus are detected only from $K_S$ and up to radio bands (cf. text).
\label{figKs}}
\end{center}
\end{figure*}


We observed He\,2-10 in $K_S$ using ISAAC at the ESO VLT/Antu 
under 0.3$\arcsec\,$ seeing (Figure~\ref{figKs}), and in $L'$ and $M'$ with 
the adaptive optics system NAOS-CONICA at the ESO VLT/Yepun, reaching a 
corrected PSF full width at half maximum (FWHM) of 0.12$\arcsec$. 
The observations were obtained in service mode in 2003-2004.
We used standard observing strategies. In $L'$, we alternated exposures
 between the object and the sky in 15 ABBA sequences with a throw of 40\arcsec.
Object cubes (respectively sky cubes) consisted of 220 (110) random 0.175\,s 
jitter for a total exposure time of 1155\,s (577.5\,s). In $M'$, as NACO was 
still in early operation, we used the only allowed chopping mode with 
pre-defined exposure time and a 30\arcsec chopping throw. Typical strehl 
ratios of validated exposures were in the range 20-25\% for both filters.

The ABBA sequences were combined/substracted using custom iraf scripts.
We performed accurate astrometry reconstruction and photometry as described 
in the following sections.

\subsection{Astrometry}\label{astrometry} 

In order to compare the astrometry of our NAOS-CONICA images with
previous multiwaveband observations, we performed an accurate calibration
of the NAOS-CONICA to the USNO-B based J2000 epoch. 
We first calibrated the astrometry of the $K_S$ image using 23 field
stars  catalogued in the USNO-B \citep{monet98}, and applied general
transformations using the NOAO iraf routine geomap. 
The positional accuracies of the transformations are 
$\Delta\alpha_{rms} = 0.008$\,sec and $\Delta\delta_{rms} = 0.08\arcsec$. 
We then derived the transformations between 
the $K_S$ USNO-calibrated image and $L$ image using four common
sources in the central region (two field stars and two bright clusters). 
The number of available common sources is too small to derive a robust 
general transformation but the clear morphological similarities between 
the $K_S$ and $L$ images (Figure~\ref{figmulti}) make us confident that 
the reconstructed $L$ astrometry has an accuracy equivalent to that of 
the $K_S$ image. 
The positional accuracies of the transformations are 
$\Delta\alpha_{rms} = 0.001$\,sec and $\Delta\delta_{rms} = 0.012\arcsec$. 
The resulting astrometry, computed for all $L'$ sources, is shown 
Table~\ref{tabastro}.

\begin{deluxetable}{lcclcc}
\tablewidth{0pt}
\tablecaption{He\,2-10: astrometry of the $L'$ bright sources of the nucleus (Fig.~\ref{figLp})\label{tabastro}}
\tablehead{\colhead{Name\tablenotemark{a}}&\colhead{R.A. (J2000)}&\colhead{Dec. (J2000)}&\colhead{Name\tablenotemark{b}}&\colhead{R.A. (J2000)}&\colhead{Dec. (J2000)}\\
\colhead{Band $L'$}&\colhead{h~min~sec}&\colhead{$^\circ~\arcmin~\arcsec$}&\colhead{Band 2\,cm}&\colhead{h~min~sec}&\colhead{$^\circ~\arcmin~\arcsec$}\\
&\colhead{$\pm$0.003sec}&\colhead{$\pm$0.02sec}&&\colhead{$\pm$0.007sec}&\colhead{$\pm$0.1sec}}
\startdata
L1 & 8~36~15.032 & -26~24~33.90 & knot 1 &8~36~15.014 &-26~24~33.81\\
L2 & 8~36~15.066 & -26~24~34.03 & knot 2 &8~36~15.060 &-26~24~33.98\\
L3a & 8~36~15.131 & -26~24~34.24 & knot 3 &8~36~15.127 &-26~24~34.13\\
L3b & 8~36~15.153 & -26~24~34.10& knot 3 &8~36~15.127 &-26~24~34.13\\
L4a & 8~36~15.218 & -26~24~34.00& knot 4 &8~36~15.234&-26~24~34.00\\
L4b & 8~36~15.247 & -26~24~34.35& knot 4 &8~36~15.234&-26~24~34.00\\
L4c & 8~36~15.251 & -26~24~34.83& knot 4 &8~36~15.234&-26~24~34.00\\
L4d & 8~36~15.265 & -26~24~34.39& knot 4 &8~36~15.234&-26~24~34.00\\
L5 & 8~36~15.310 & -26~24~35.11& knot 5 &8~36~15.308&-26~24~34.61\\
L6 & 8~36~15.183 & -26~24~34.06&&&\\
L7 & 8~36~15.191 & -26~24~34.60&&&\\
L8 & 8~36~15.217 & -26~24~34.50&&&\\
\enddata
\tablenotetext{a}{FWHM $0.12\arcsec$ (This work).}
\tablenotetext{b}{Beam size $\sim0.4\arcsec\times0.8\arcsec$ (\citet[Table~3]{kobulnicky99}). The astrometry is from \citet{kobulnicky99}.} 
\end{deluxetable}

\subsection{Photometry}\label{photometry}

\begin{figure*}
\begin{center}
\includegraphics[height=5.7cm]{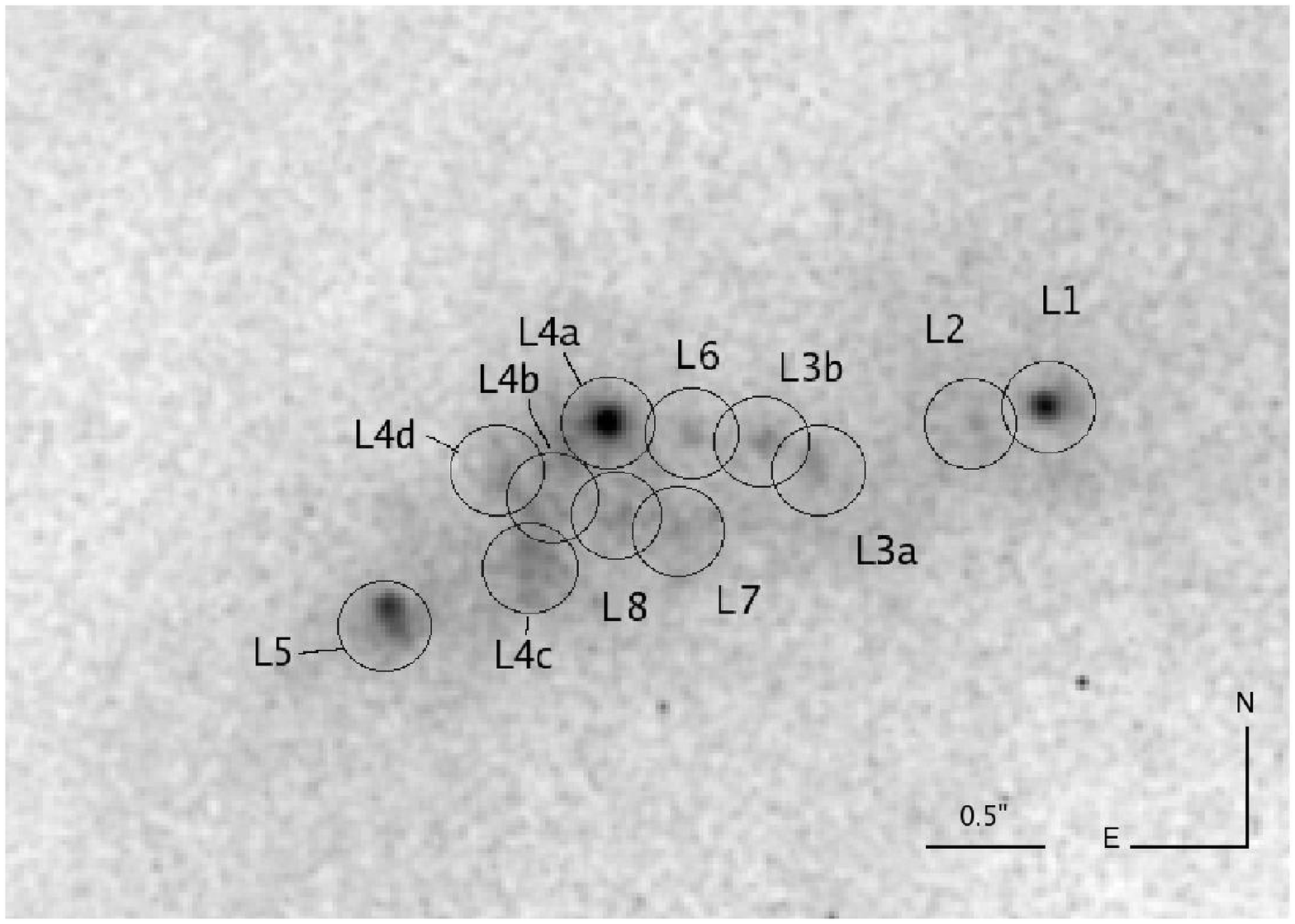}
\hfill
\includegraphics[height=5.7cm]{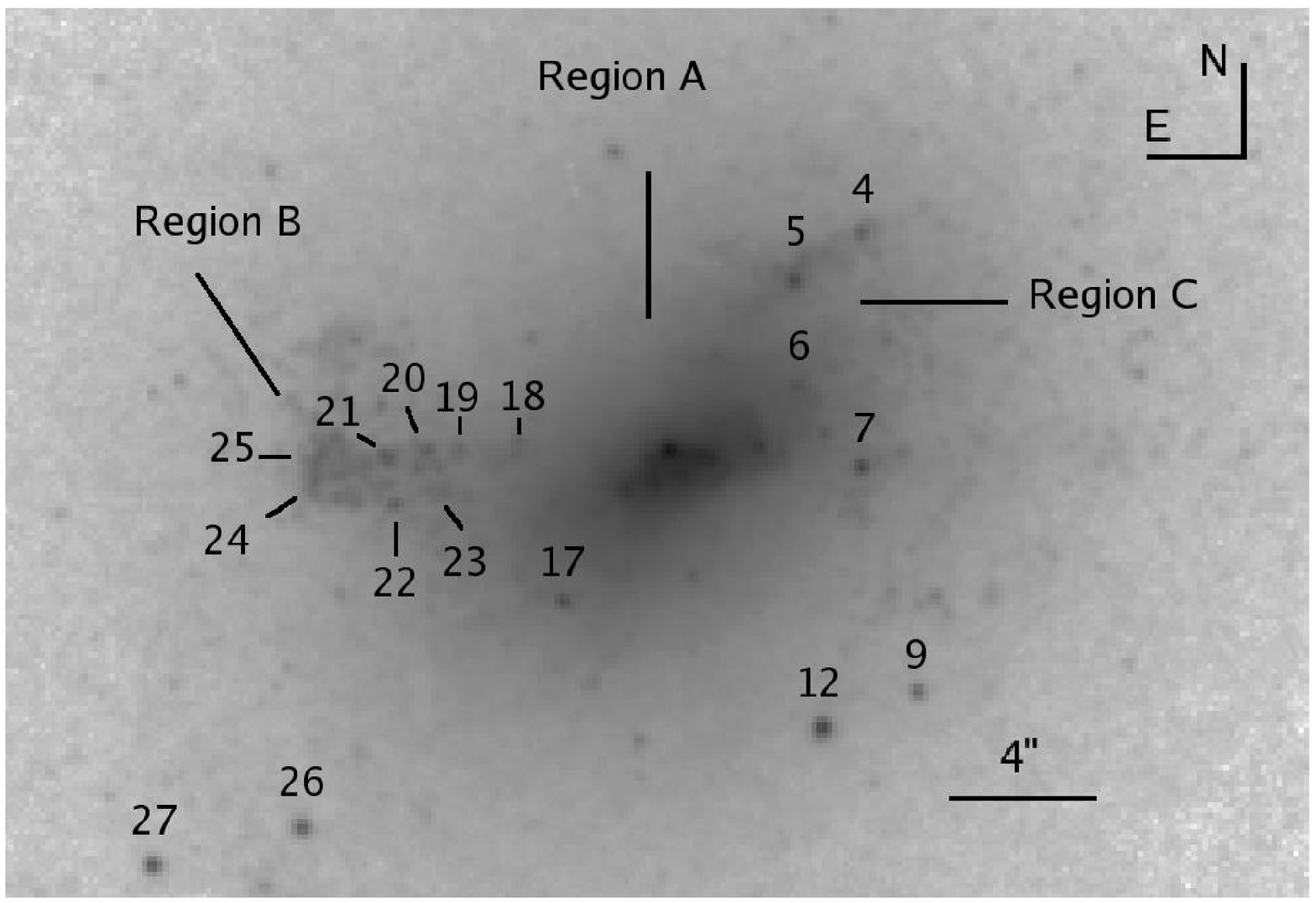}
\caption{\footnotesize Left: He\,2-10 nucleus is shown in $L'$, 
with VLT/NAOS-CONICA, FWHM 0.12 arcsec, Field of view $\sim7\arcsec$, with labels referring to the source naming convention used in this paper.
The apertures are $0.5\arcsec$~wide, or $22.5\,$pc at $9\,$Mpc.
Right: He\,2-10 is shown in Ks + finding chart of
outer source with $M_V<-6.5$ detected in $V$, $I$,\& $K_S$. Source 18 to 25 
belong to region~B, sources 4 to 7 to region~C.\label{figLp}}
\end{center}
\end{figure*}

We extracted $L'$- and $M'$ fluxes 
of all detected sources of the inner core of He\,2-10 in aperture sizes 
of 0.5$\arcsec$ (the finding chart in $L'$ is given Figure~\ref{figLp}, left panel), 
the good match between VLT-NACO and HST 
resolutions allows us to compute consistent aperture photometry 
over the entire optical and near-infrared (NIR) range.
We measured the $F555W$, H$\alpha$, $F814W$ fluxes from the
HST calibrated archive images, on the same positions and aperture sizes, 
including a systematic centroid position mismatch in the error computation. 
All photometric measurements were then calibrated to Vega magnitudes 
using IR standards provided by ESO service observing team. All ancillary
data may be found in \anchor{http://archive.eso.org/eso/eso\_archive\_main.html}{http://archive.eso.org/eso/eso\_archive\_main.html},
under Program ID 71.B-0492(A). Typically, 
$L'$ errors on zeropoints are ca. 10\%, because of a highly variable 
background. Additional uncertainties are due to the anisoplanetism of 
the FOV of view and the different strehl ratio between the photometric
standard and the science observations. Both effect are difficult to quantify,
but we assume that, because of our large apertures ($5\times$FWHM), 
the strehl difference will not dominate the systematics
and most of the error will actually come from background subtraction.
Unfortunately NACO's response was not fully known when the $M'$ 
data were taken and the standard stars were observed with high ADU counts.
They possibly reach the non-linear regime of the detector and the derived 
zero-points from different nights disagree to $\pm0.25$\,mag. 
Hence the $M'$ data are mostly useful for morphology and only deliver a crude 
photometry.

HST $F555W$, $F658N$/H$\alpha$, and $F814W$ images were analysed following 
standard recipes detailed by \citet{holtzman95}. We extracted the sources of 
the nucleus
with the full-resolution images in order to compare with $L'$ data, and
we convolved the HST data with a $0.3\arcsec$ gaussian in order to compare
with $K_S$ data.
We derived H$\alpha$ equivalent width interpolating the continuum between
$F555W$ and $F814W$ following \citet{johnson00}.
We recover the original magnitudes and similar H$\alpha$ equivalent widths 
of \citet{johnson00} and the V magnitude of knot 4 \citep{chandar03} using
different strategies with smaller apertures within photon and aperture 
position errors. 

All Vega calibrated magnitudes were corrected for the 
galactic extinction of $E(B-V)=0.11$ \citep{schlegel98} yielding 
$A_V=0.369$, $A_{\mbox{H}\alpha}=0.298$ $A_I=0.216$, $A_K=0.04$, $A_L=0.017$, 
and $A_M=0.0$. 

We also extracted all sources detected in $K_S$, $F814W$ and $F555W$ 
in the outer 1\arcmin\,of He\,2-10 (0.5$\arcsec\,$ aperture size, 
see Figure~\ref{figLp} right panel). 
A special care was taken to check the effect of the larger PSF in $K_S$ 
and HST convolved resolution, compared to HST original resolution. 
We computed the total flux lost between unconvolved and 
convolved point sources for 0.5$\arcsec\,$ apertures. 
The loss amounts to $0.31\pm0.05$ mag in $F814W$ and $F555W$. 
Although substantial in absolute, the correction is negligible in color, 
hence the color-color diagrams are not affected by
the correction applied to the sources of the outer 1\arcmin. 

Finally, another photometric extraction was done on the brightest $L'$ 
and $M'$ sources in order to build the full spectral energy distributions 
of these 
regions.
We convolved $L'$ and $M'$ images with 0.3\arcsec~and extracted fluxes 
around the brightest radio knots in 1\arcsec~apertures.

The largest source of uncertainty comes from the diffuse background in the 
inner core of He\,2-10. We adopted different strategies to estimate the 
background in the different bands, in order to account for the variable 
resolutions and field crowding of the data.
The HST images were sky subtracted using sky annuli of radius 
0.3-0.5$\arcsec\,$ around the objects located in the inner 10$\arcsec\,$ of
the nucleus, and using annuli 0.4-0.6 pix for outer sources (also for $K_S$). 
In $K_S$, $L'$ and $M'$, we computed the diffuse emission background in annuli 
of radius 0.3-1$\arcsec$ around the objects located in the 
inner 10$\arcsec\,$ of the nucleus. We additionally computed
2 sky annuli (0.3-0.5$\arcsec$, and 0.3-2$\arcsec$) in order to estimate
the systematic errors resulting from the measured flux differences. 
Sytematic errors of $0.15$ mag were typical. The $M'$ magnitudes 
have higher systematics due the uncertainty on the zeropoints. 
Finally the background around the 1\arcsec~apertures was measured
in annuli of 0.5-1.5\arcsec.

\begin{deluxetable}{lccccc}
\tablewidth{0pt}
\tablecaption{0.5\arcsec\,aperture photometry of He\,2-10 (Fig.~\ref{figLp})
\label{tabcmd}}
\tablehead{\colhead{Source \#}&\colhead{$F555W$}&\colhead{$F814W$} &\colhead{$K_S$}&\colhead{$L'$}&\colhead{$\log($EW[H$\alpha$])}\\
&\colhead{Vega}&\colhead{Vega}&\colhead{Vega}&\colhead{Vega}&\colhead{log(\AA)}}
\startdata
L1 & 19.48$\pm$0.08 & 18.53$\pm$0.06 & 15.61$\pm$0.03 & 13.53$\pm$0.01 & 2.60\\
L2 & 19.21$\pm$0.05 & 18.36$\pm$0.04 & 15.80$\pm$0.02 & 14.09$\pm$0.90 & 2.66\\
L3a& 16.94$\pm$0.07 & 16.53$\pm$0.07 & 14.79$\pm$0.08 & 14.07$\pm$0.01 & 1.39\\
L3b& 16.61$\pm$0.03 & 16.19$\pm$0.02 & 14.53$\pm$0.03 & 14.08$\pm$0.03 & 0.74\\
L4a& 16.16$\pm$0.04 & 15.77$\pm$0.03 & 13.92$\pm$0.03 & 13.04$\pm$0.04 & 1.69\\
L4b& 16.68$\pm$0.02 & 16.38$\pm$0.01 & 14.65$\pm$0.02 & 13.49$\pm$0.01 & 2.32\\
L4c& 16.71$\pm$0.05 & 16.38$\pm$0.03 & 14.76$\pm$0.04 & 13.48$\pm$0.02 & 2.39\\
L4d& 17.62$\pm$0.17 & 17.06$\pm$0.13 & 15.06$\pm$0.13 & 13.78$\pm$0.08 & 2.38\\
L5 & 18.99$\pm$0.17 & 18.14$\pm$0.14 & 15.29$\pm$0.06 & 13.34$\pm$0.02 & 2.37\\
L6 & 16.65$\pm$0.02 & 16.30$\pm$0.02 & 14.55$\pm$0.02 & 14.10$\pm$0.06 & 1.23\\
L7 & 17.67$\pm$0.04 & 17.04$\pm$0.03 & 14.75$\pm$0.02 & 13.83$\pm$0.02 & 2.21\\
L8 & 17.26$\pm$0.03 & 16.77$\pm$0.03 & 14.64$\pm$0.02 & 13.68$\pm$0.02 & 2.28\\
\enddata
\end{deluxetable}

\section{Results}\label{results}

\subsection{Optical to radio identification of the central sources}
\label{secident}
\begin{figure*}
\begin{center}
\plotone{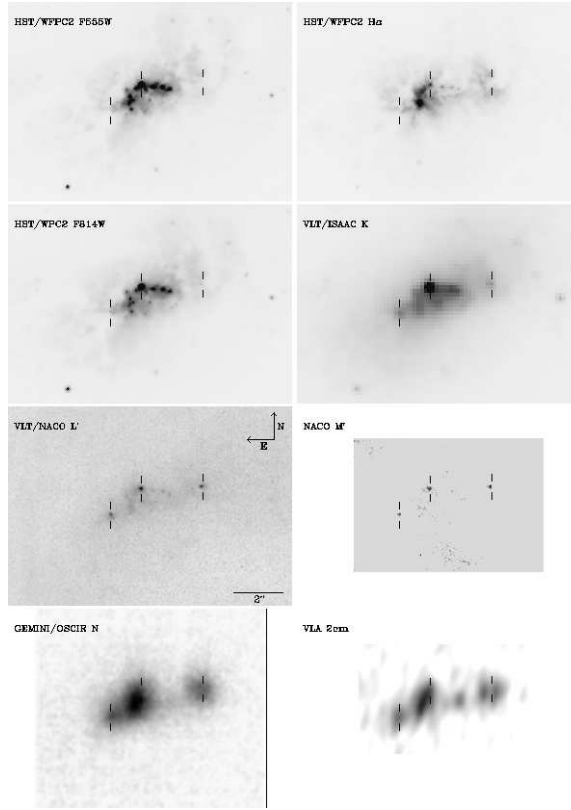}
\caption{\footnotesize The nucleus of He\,2-10 is shown from blue to radio
wavelengths, from top-left to bottom-right. Each panel shows the central 
region observed with VLT/NAOS-CONICA in $L'$-band. Three bright regions 
are gradually emerging from the dust from $K_S$ to radio bands. {All images are now plotted on a common, identical, astrometric grid, the vertical markers replicate the position of the brightest $L'$ sources in all panels.}\label{figmulti}}
\end{center}
\end{figure*}

\begin{figure*}
\begin{center}
\includegraphics[width=8cm]{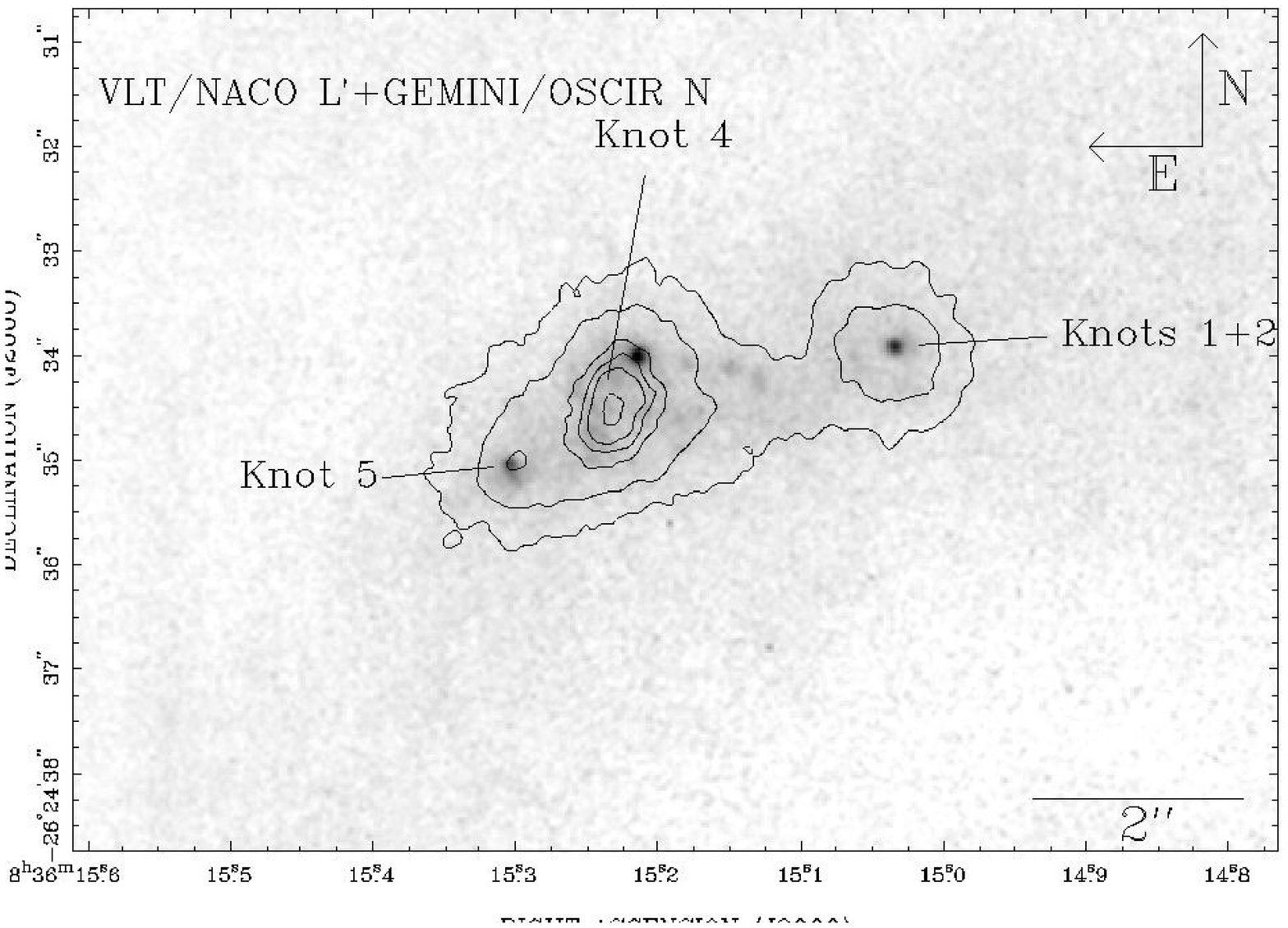} 
\includegraphics[width=8cm]{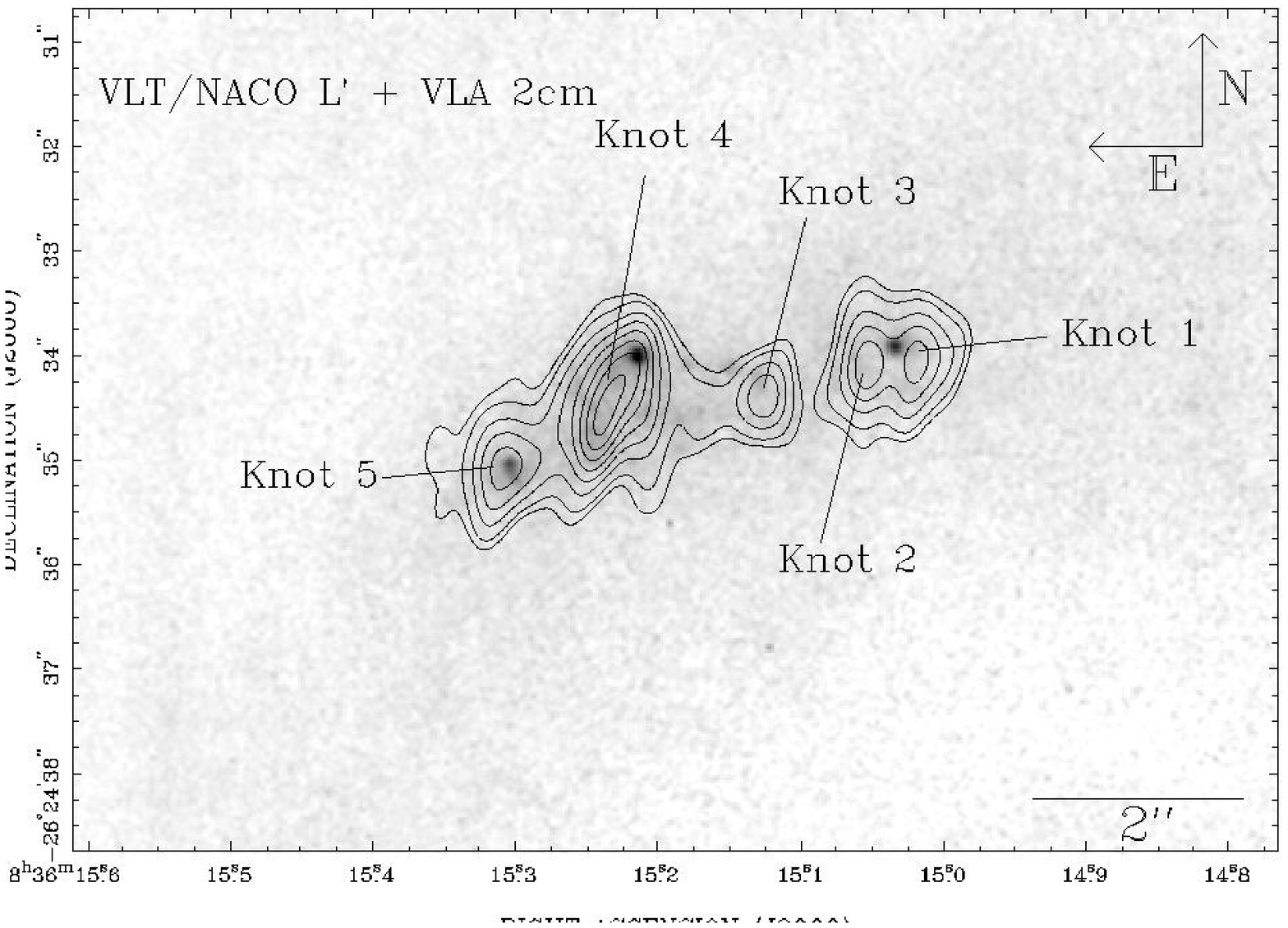}\\
\caption{\footnotesize Left: $L'$ gray-scale image overplotted 
with Gemini $N$ contours. Right: $L'$ gray-scale image overplotted 
with VLA 2-cm contours. 
The $N$ sources are named using the identification of \citet{vacca02}. 
The radio knots are named according to \citet{kobulnicky99}. 
The brightest $L'$ sources are named following the radio knots (cf. text 
and Fig.~\ref{figLp}).\label{figcontour}
}
\end{center}
\end{figure*}

\begin{figure}
\begin{center}
\plotone{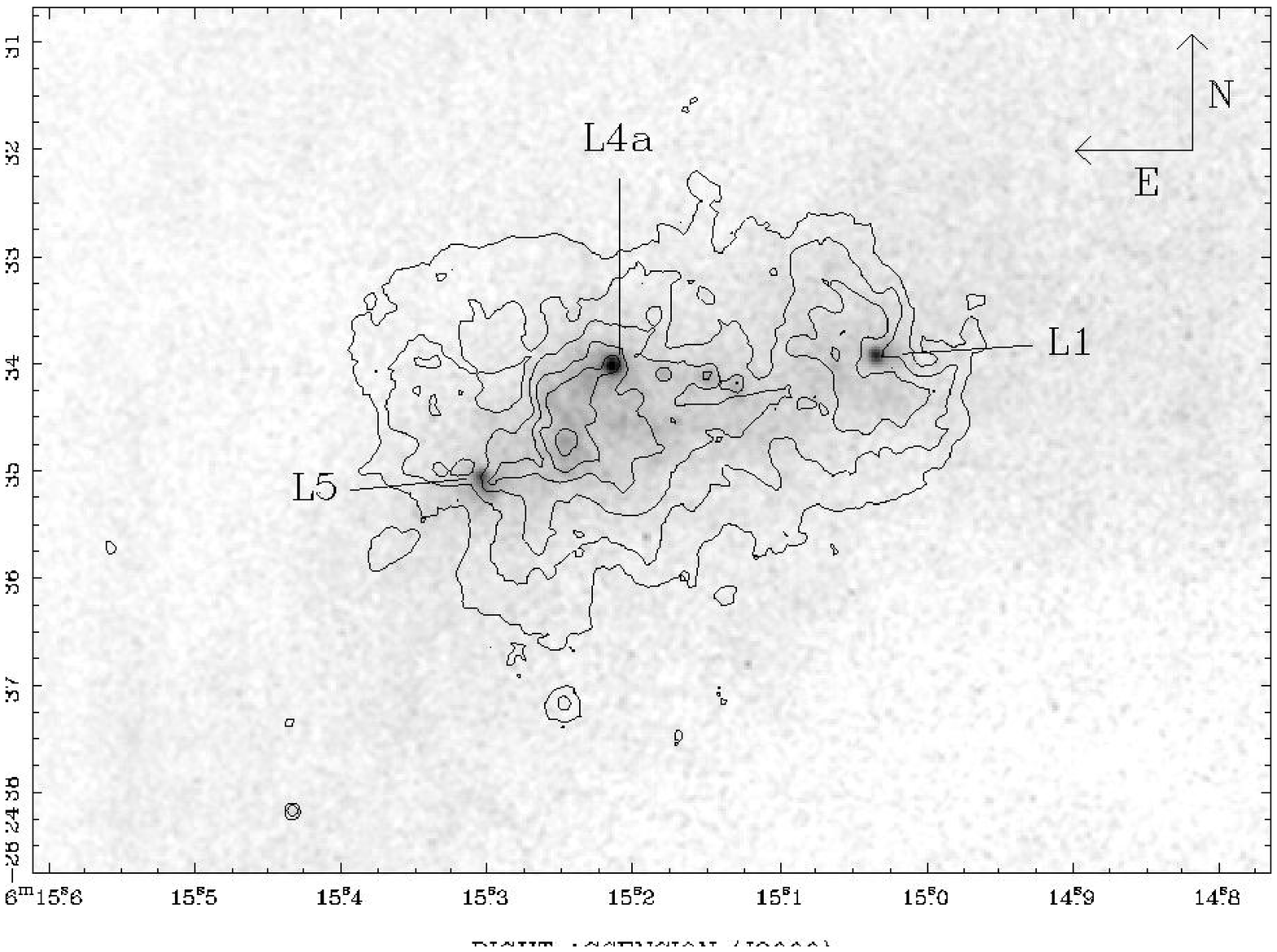}
\caption{\footnotesize NACO $L'$ gray-scale image overplotted 
with HST $F658N/H\alpha$ contours.\label{figHa}}
\end{center}
\end{figure}

Figure~\ref{figmulti} shows the entire series of
high-resolution observations of the nucleus of Henize 2-10,
in $F555W$, $F658N/H\alpha$, $F814W$, $K_S$, $L'$,
$M'$ (shifted to USNO-B absolute astrometry, see 
section \ref{astrometry}), $N$ and $2$\,cm. The $M'$ image is much shallower 
than the $L'$ image, and only the three reddest sources are robustly detected. 
As is evident from this figure, and also from Table~\ref{tabastro}, the $L'$ 
and $M'$ sources are clear counterparts to some of the radio knots, and
because of the strong morphological and physical association between the radio 
and MIR emissions, to the MIR sources as well.  
Figure~\ref{figcontour} shows a precise comparison
of the $L'$ images (gray scale) with $N$ GEMINI/OSCIR $N$ contours 
\citep[left, the $N$ source associations to the radio knots are
according to ][]{vacca02}, and with VLA 2-cm contours \citep[right, with the 
radio knots identification of ][]{kobulnicky99}. It is striking to see, on 
Fig~\ref{figcontour}, how well sources L1 and L5 correlate with the N-band 
sources. 
Furthermore Figure~\ref{figmulti} also demonstrates the association of L4a 
with the brightest source in F814W, and of L6, L3a and L3b with the string 
of three clusters detected west of the brightest source in F814W. Therefore, 
in constrast to the statement of \citet{kobulnicky99,kobulnicky00}, 
we see a clear connection between the structures observed from 0.5\,$\mu$m 
to 2\,cm. 

At this stage it is important to stress that the positional association 
from the optical to the radio that we present here is not a new proposition 
that we are making, but is a fact imposed by the astrometry. The HST, 
$K_S$, $L'$, and $M'$ images are now tied to the same astrometric reference, 
which is the USNO-B astrometry. This is accurate to $\pm0.3\arcsec$ which 
will be the dominant positional uncertainty in this wavelength regime 
(for instance the relative astrometric accuracy of NAOS-CONICA is 
$\pm0.02\arcsec$). The absolute astrometry of the VLA is accurate to 
$\pm0.1\arcsec$. With these accuracies, it becomes impossible to accept 
the relative positions of the HST and MIR/radio sources as presented 
in \citet{kobulnicky99,kobulnicky00,vacca02}, i.e. a shift of 1.2\arcsec\ 
in $\alpha$, even taking into account the beam size of the VLA, which is 
0.4\arcsec\ in $\alpha$. In fact, the main source of astrometric error in 
previous papers was an incorrect astrometry in the HST data used at the time. 
This is now corrected in the archived versions of the same data. Therefore 
Figure~\ref{figmulti} now reveals for the first time the correct evolution 
of the central region of He\,2-10 from the optical to the radio. This allows 
now to present a much firmer identification of the different components 
across the spectrum.

L1 is clearly detected in $M'$ and can now be associated to 
the western N band source. It has a rather faint counterpart in $K_S$ 
and is no longer detected in the optical bands, and shows a clear
anti-correlation with H$\alpha$ (Fig.~\ref{figHa}). 
Figure~\ref{figcontour} shows that L1 appears to fall between the radio 
knots 1 and 2, however this time the shift is too small to be significant. 
Since L1 is accompanied by a faint source L2, we propose to associate the 
two sources L1 and L2 with knot\,1 and knot\,2 in the radio map. L2 has 
no counterpart in the optical bands, an anti-correlation with H$\alpha$ 
(Fig.~\ref{figHa}), much like L1, but is much fainter. Source L3a and L3b 
have clear optical and H$\alpha$ counterparts west of the brightest optical 
source (cf. Fig.~\ref{figHa}). They fall in a region of diffuse N-band 
emission and their association with knot\,3 is unclear. 
The brightest $L'$ source L4a is the counterpart of the main source 
in the F814W image. It is fainter in the H$\alpha$ image but reappears 
strongly in the F555W image. Again L4a appears displaced from its possible 
radio and MIR counterpart knot\,4. In fact it is quite likely that knot\,4 
and the associated brightest MIR source are the counterparts of the group L4b, 
c and d. These sources are easily detectable in the optical wide bands and 
are quite strong in H$\alpha$. L5 is another bright source that has no 
optical or H$\alpha$ counterpart (Fig.~\ref{figHa}) but is cleary detected 
in $K_S$, N and radio and can be associated with knot\,5. L6 has no clear 
radio or MIR counterpart but is associated to the optical cluster just west 
of the brightest F814W source. Finally L7 and L8 have no definite counterparts 
but are associated with diffuse emission in all bands.

We emphazise that the fact that we detect spatially correlated sources 
accross the spectrum doesn't mean that we actually detect the \emph{same} 
sources.  An error of 0.2\arcsec\ translates to 9\,pc at He\,2-10's distance, 
while the compact sources are not expected to have sizes much larger
than 1-2pc, hence the $L'$ sources and radio knots could belong 
to contiguous yet distinct star forming regions. 
A robust identification will have to wait for deeper $L'$ observations 
and higher resolution radio observations. Nevertheless, it is interesting 
to follow the working hypothesis that association between radio and NIR 
sources is indeed a physical one.

\subsection{Near-infrared/optical colors of He\,2-10 nucleus}

\begin{figure}
\plotone{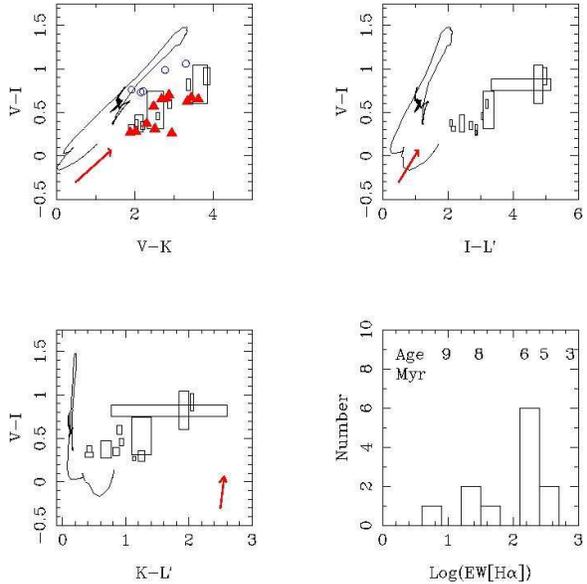}
\caption{\footnotesize Color-color diagrams of all 12 sources detected in $L'$ 
with VLT-NACO in Henize 2-10 central nucleus (squares; sizes proportional 
to errors), and of 21 sources visible in the VLT/ISAAC $K_S$ covering
the outer 1\arcmin\, around the nucleus (open circles are for the most
remote sources, filled triangles are sources surrounding the nucleus, mostly
from so-called regions B and C). The outer sources are visible only up to 
$K_S$ hence they are present only in the top-left diagram. Only sources
brighter than $M_V<-8.5$ are included. 
V and I magnitudes were extracted from HST archive images (cf. text). 
The thin solid lines are dust free STARBURST99 track of solar metallicity, 
Salpeter IMF of instantaneous bursts from 0-1Gyr (low $V-I$ to high $V-I$). 
The thick arrow shows the amplitude and direction of $A_V=1$\,mag internal 
extinction. Bottom graph shows the H$\alpha$ equivalent width 
(Log(EW[H$\alpha$]) of the 12 $L'$ sources computed from an HST $F658N$ 
archive image (cf. text).
\label{figcmd}}
\end{figure}

We selected the sources likely to be clusters at 9~Mpc with a magnitude 
threshold of $M_V<-8.5$ \citep{whitmore99,johnson00}. Table \ref{tabcmd}
shows the corresponding Vega magnitudes of the 12 central sources 
(Fig~\ref{figLp}). 
All 12 sources seen in $L'$ would still be too bright to be supergiant 
stars if Henize 2-10 was at a distance of 6~Mpc, but 2 sources out of the 21 
outer sources detected in $V$, $I$ and $K_S$ would fall below the threshold. 
We plot $V-I$ vs $V-K$,  $V-I$ vs $I-L'$ and $V-I$ vs $K-L'$ colors in 
Figure~\ref{figcmd} along with the dust free STARBURST99 
\citep[SB99]{leitherer99} tracks of solar metallicity instantaneous bursts, 
with Salpeter IMF,  from 0 (low $V-I$) to 1\,Gyr (high $V-I$). 
The sharp turnoff seen in $V-I$ vs $K-L'$ 
correspond to 6.3\,Myr,  after which nebular emission ceases to be important. 
The 12 sources of the nucleus are plotted as squares the size of which relate 
to the total errors. The source of the outer 1\arcmin are plotted as filled
triangles for the sources belonging to the regions B and C of the north-east
and west/north-west (Fig.~\ref{figKs}), and as circles for sources falling 
at least 1\arcmin\,away from the nucleus (all circles happen to be in the 
southern part of He\,2-10, visible as bright blue [probably foreground] 
clusters in Fig.~\ref{figKs}). Photometric errors are typically contained 
within the size of the symbol, but systematics related to the diffuse 
background subtraction can be of the order of 15\%.

The solid arrows show the direction and the amplitude of a
screen extinction of $A_V=1$ mag (assuming a Calzetti form instead of the
canonical galactic extinction curve does not change the values by more than
3\% in the range 0.55\,$\mu$m-3.8\,$\mu$m). 
All diagrams show the well-known age-extinction degeneracy of optical 
colors (here $V-I$), as well as the gradual decrease of the extinction 
impact on colors as one goes to the infrared.  All show that most of the 
data points cannot be explained by reddenning any point of the models 
by any amount of screen extinction, especially for the diagrams including 
the $L'$ photometry. This is a rather common observation in starburst regions
\citep{johnson04,vanzi04,cresci05}.

On the top left panel, the squares and the triangles seem to occupy the
same locus whereas the circles are shifted up by $V-I\simeq0.3\pm0.1$\,mag.
This shift can be due to a systematic effect in the background subtraction
although unlikely because the colors of triangles and squares were derived 
with independent methods (cf section \ref{photometry}) and still lies on top
of each other, hence do not seem to be dominated by systematics. 
The shift more likely comes from an infrared excess affecting the central 
sources of He\,2-10.
This conclusion finds some support in the large excesses of
$I-L'$ and $K-L'$ colors observed for the 12 central sources in the other 
two color-color diagrams. 
These infrared color excesses can have a number of explanations that will 
be explored later. But let us first characterize 
the magnitude of the extinction affecting the central sources. This will 
help us setting constraints on clusters ages as well as on their actual 
infrared excess from their location in the color-color diagrams.


As expected in a complex region such as revealed by Figure~\ref{figKs}, 
extinctions measured at different wavelengths do not agree well with one 
another. First \citet{allen76b} measure $A_V$=2.3 from optical 
observations while \citet{johansson87} correcting for the contribution of 
stellar absorption features obtain $A_V=0.86$. Then we can use the observation 
of \citet{vanzi97} to derive the extinction of the central part of He\,2-10 
from the H$\alpha$ to Br$\gamma$ ratio. 
\citeauthor{vanzi97} measure a Br$\gamma$ flux of 
6.3$\times10^{-14}$\,erg$\cdot$s$^{-1}\cdot$cm$^{-2}$ over an aperture of 
2.4x15.6\arcsec centered on the K bright nucleus. From the HST H$\alpha$ 
image we measure the H$\alpha$ flux on the same aperture and obtain a value 
of 1.86e-12 erg/s/cm2. Assuming an intrisic H$\alpha$/Br$\gamma$ ratio of 
155.4 corresponding  to Te=7500\,K, ne=1000cm$^{-3}$ \citep{storey95} 
we find $A_V = 1.25$.
Finally, from the ratio Br$\gamma$/Br10 of Vanzi \& Rieke we obtain 
$A_V$=10.5 which is in full agreement with the extinction derived using 
the Br$\alpha$, Br$\gamma$ fluxes of \citet{kawara89}. We thus observe a 
clear trend of increasing extinction from the optical to the IR which is 
typical of a system where the absorbing material, gas and dust is mixed 
with emitting sources. It is worthwhile to note here that correcting the 
clusters for extinction will bring them closer to the youngest part of 
the SB99.

But this correction cannot explain the significant red excess observed 
in $V-K$, $I-L'$ and $K-L'$ for all the central sources. This red excess 
was also observed in a similar diagram built for NGC\,5253 
\citep{vanzi04,cresci05} and for Haro\,3 \citep{johnson04}, and most likely 
has the same origin: a contribution of hot dust in the NIR bands. 
Indeed as the earliest evolutionary phase 
of star clusters occurs deep in molecular clouds, we can expect to find dust 
and molecular clouds in the immediate vicinity of these clusters. Dust close 
to its sublimation temperature would be able to contaminate the $L'$ band. 
Using radiative transfer models, \citet{vanzi04} showed that 
the location of the reddest clusters in the ($V-I$,$K-L'$) diagram of 
NGC\,5253, could be explained by a combination of extinction, scattering and 
emission by dust in a shell around the cluster. This is likely what we 
observe here as well. In that respect, it is noteworthy that the three 
reddest sources in $K-L'$ are L1, L2 and L5, the sources with no visible 
counterparts and associated with thermal radio knots. If we apply to the 
12 central sources of He\,2-10 color corrections such as those derived 
from the modeling of the red cluster in NGC\,5253, this will bring all 
the sources to the bottom left of the diagrams, indicating ages of less 
than 6.3\,Myr.

\subsection{Physical properties}

We study now the physical properties of the sources which dominates the
$L'$ emission of the nucleus (Fig.~\ref{figLp}).
The bright $L'$ sources L1+L2, L4 and L5 are not or barely resolved 
with radii $\sim0.1\arcsec$ or 4.5~pc, L5 is not isolated, 
there is a faint diffuse emission extending as far as 3 FWHMs (10~pc)
(the apertures shown on Fig.~\ref{figLp} are 22.5pc wide). These sources
are not resolved in $M'$ either (radii$\la0.12\arcsec$).
Fig.\ref{figcmd} bottom panel shows the histogram of the H$\alpha$ 
equivalent width Log(EW[H$\alpha$]) of the 12 central sources in units 
of Log[\AA] (from Table \ref{tabcmd}) and the associated ages 
according to \citet{leitherer95}. The histogram peaks at ages $\la6$~Myr in 
agreement with recent works \citep{johnson00,chandar03} and with age trends
inferred from previous section color analysis. We emphasize that the 
H$\alpha$ equivalent width measurements are prone to strong systematic 
effects. We substracted out the [NII] $\lambda 6584$ measured to be 31\% 
of H$\alpha$ in an unpublished high-resolution spectrum covering the 
1.6\arcsec~around source $L4a$. 
But the main source of error is a poorly-known continuum level. Our method to
derive the continuum by interpolation of the contiguous broadband filters 
continuum for lack of a better estimator is crude and the 
quoted Log(EW[H$\alpha$]) should be carefully used.

The measured Log(EW[H$\alpha$]) centered on the sources L1, L2 
and L5 are 2.60, 2.66 and 2.37, yielding ages$\la5$\,Myr.
But the H$\alpha$ emission shows remarkable anti-correlations with the 
source L1+2 and L5 and no emission in $F555W$, $F658N$ and $F814W$ can
be clearly associated with source L1, L2 and L5. Hence the measured 
Log(EW[H$\alpha$]) is probably more related to He\,2-10 foreground 
medium rather than the $L'$ emitting clusters. 
An interesting check on the age derived from Log(EW[H$\alpha$]) 
comes from the sources L3a, L3b, L4a and L6 because all three have 
unambiguous $F555W$, $F658N$ and $F814W$ counterparts. 
Log(EW[H$\alpha$])$<1.5$ and the brightest source in $L'$ 
(L4a of Fig.~2) have Log(EW[H$\alpha$])$<1.5$ (Table \ref{tabcmd}), 
consistent with the measurements of \citet[Fig.~9]{johnson03}, yielding 
ages$>$6\,Myr, whereas \citet{chandar03} derived ages of 4-5\,Myr fitting 
single-burst SB99 templates to a de-reddened UV/STIS spectrum of the same 
sources. Thus there appears to be an age dichotomy, with source L1, L2 
and L5 being younger than 5\,Myr and the other $L'$ source being older.

On the basis of the new identification of the components accross the 
spectrum (see Section~\ref{secident}), it is important to review previous 
assumptions on the nature of He\,2-10 radio knots emission.
The presence of UD HII in He\,2-10 is primarily inferred from observed 
turnovers from flat radio spectra ($F_\nu = \nu^{-\alpha}$ with $\alpha=0$,
characteristic of an optically thin ionized gas), to thermal
spectra ($\alpha=2$, signature of a free-free optically thick gas)
\citep{kobulnicky99}. This feature is common in galactic HII regions. 
These UD HII regions were presumed to be extremely young, e.g. less than one 
million year, mostly because of their compactness, their high electron 
densities, and of their lack of counterpart in the NIR and below. Indeed 
only HII regions buried extremely deep in dust, i.e. right at the start of 
their expansion, would remain invisible in the NIR. Since most of the radio 
sources now have counterparts, the interpretation of the radio knots must be 
updated.

Among the 5 radio sources, we would only classify knot 1+2 and knot 5, as
bona fide UD HII regions. They have counterparts in the NIR but not in the 
visible. This implies a significant optical depth (typically $\ga$\,10 if 
we follow existing models of embedded super-star clusters), and thus a young 
age although possibly not as young as previously postulated.


The radio knot 3 was already known to be a different type of source than 
other radio knots from a strong non-thermal signature \citet{johnson03},
(very likely a supernova remnant). We also propose that the radio knot 4, 
associated with diffuse $L'$ emission (L4b,c,d) and showing strong 
correlations with bright  optical and H$\alpha$ sources, is not an UD HII 
region but a complex mix of {\em normal} HII regions and supernova remnants. 
Knot 4 does show a non-thermal signature with a slightly
negative spectral index $\alpha$.
The proposed re-classification of the radio sources would lower the 
total mass of hidden $O$ stars (as deduced from the radio) by a factor 
of $\sim$2 \citet{johnson03}. 

Detailed modelling of the spectral energy distributions is essential
to have a deeper understanding of the star formation process ongoing
in He\,2-10, especially of sources L1+L2 and L5 which represent the best 
candidate for very young dust-enshrouded super-star clusters. 
This is however a long endeavour that we differ to a later paper. 
Nevertheless the present observations outline the importance of
high-resolution multiwavelength datasets to disentangle the intrinsically 
complex nature of star forming regions.  

\section{Conclusions}\label{Conclusion}

We detected compact sources in the nucleus of Henize 2-10 
with $K_S$, $L'$ and $M'$ observations using ISAAC and NAOS-CONICA on the VLT. 
The sources are compact ($<4.5$~pc), highly correlated with 
radio and mid-infrared ultradense HII regions, previously thought 
to be optically thick.

The color-color magnitude diagrams show the presence of strong
red-excess in $K_S$ and $L'$. Such red excesses point at highly
heterogeneous dust distribution and at the presence
of a hot dust component emitting and scattering down to $L'$ and $M'$.

We tentatively review the previous classification of the radio
knots by identifying two bona fide UD HII (knot 1+2 and knot 5)
and propose that knots 3 and 4 are non-thermal radio sources, 
akin to supernova remnants.

These new high-resolution data uncover a complex structure in infrared.
We suggest that to understand He\,2-10 star forming history,
a detailed model of the radiative processes is needed.
This model should include all known components of the galaxy in a 
consistent way in order to fit the spectral energy distribution from 
radio to UV.

\acknowledgements
RAC wish to acknowledge an ESO grant for visiting scientist in Santiago,
and thank Chip Kobulnicky for sharing his VLA 2-cm data and Kelsey Johnson
for useful comment on the first version of the paper.\\

Facilities: \facility{VLT(ISAAC)} \facility{VLT(NAOS-CONICA)}

\end{document}